\documentstyle[12pt,epsfig]{article}
\def\Title#1{\begin{center} {\Large #1 } \end{center}}
\def\beq{\begin{equation}}
\def\eeq#1{\label{#1}\end{equation}}
\def\eeqn{\end{equation}}
\def\beqa{\begin{eqnarray}}
\def\eeqa#1{\label{#1}\end{eqnarray}}
\def\eeqan{\end{eqnarray}}
\def\CR{\nonumber \\ }
\def\leqn#1{(\ref{#1})}

\let\bar=\overbar
\def\vckm{V_{\rm CKM}}
\def\M{{\cal M}}
\def\O{{\cal O}}
\def\A{{\cal A}}
\def\hc{{\mbox{\rm h.c.}}}
\def\BR{\mbox{\rm BR}}
\def\ee{e^+e^-}

\begin{document}

\title{$B$ Decays, the Unitarity Triangle, and the Universe} 
\author{\\ Adam F.~Falk\\ \\
{\it Department of Physics and Astronomy}\\
{\it The Johns Hopkins University}\\
{\it 3400 North Charles Street}\\
{\it Baltimore, Maryland 21218 U.S.A.}}
\maketitle
\thispagestyle{empty}
\setcounter{page}{0}
\begin{abstract}
A review is given of recent developments in the physics of
flavor.  Current constraints on the Cabibbo-Kobayashi-Maskawa
matrix are discussed and related to the recent measurements of
$\epsilon'/\epsilon$, $\sin2\beta$ and
$K^+\to\pi^+\nu\bar\nu$.  A brief review is given of the
connection between CP violation in $B$ decays and electroweak
baryogenesis.  Finally, there is an extensive discussion of
how present and proposed experiments in $K$ and $B$ physics
can constrain the pattern of flavor changing processes at low
energies and, one hopes, eventually provide unambiguous
evidence of physics beyond the Standard Model.
\end{abstract}
\bigskip\bigskip\bigskip
\centerline{Plenary talk presented at the}{ XIX International
Symposium on Lepton and Photon Interactions at High Energies}
\centerline{Stanford University, August 9-14, 1999}
\vfill\eject

\Title{$B$ Decays, the Unitarity Triangle, and the Universe}
\bigskip\bigskip


\begin{raggedright}  
{\it Adam F.\ Falk\index{Falk, Adam F.}\\
Department of Physics and Astronomy\\
The Johns Hopkins University,
Baltimore, Maryland 21218}
\bigskip\bigskip
\end{raggedright}

\section{Introduction}

The past year has seen a number of exciting experimental
developments which have advanced our understanding of the
physics of flavor.  These include the confirmation by
KTeV~\cite{Alavi-Harati:1999xp} and NA48~\cite{NA48epspreps}
of the NA31 result~\cite{Barr:1993rx} for ${\rm
Re}(\epsilon'/\epsilon)$ in $K_L\to\pi\pi$, the first
measurement by CDF~\cite{Abe:1998qj} of $\sin2\beta$ in $B\to
J/\psi K_S$, and the observation of a single event in
$K\+\to\pi^+\nu\bar\nu$ by
BNL--AGS--E787~\cite{Adler:1997am}.  In addition, 1999 has
seen the startup of the next generation of $e^+e^-$
$B$ Factory experiments:  BaBar at PEP--II (SLAC), BELLE at
KEK--B (KEK), and CLEO--III at CESR (Cornell).  The first
physics results from these machines are expected in 2000. 
The fixed target experiment HERA--B, operating in the HERA
accelerator at DESY, will soon be taking physics data as
well.  For high statistics studies of the $K$ system, KLOE at
the DA$\Phi$NE $\phi$ Factory (Frascati) has also begun to
take data this year. 

Having been assigned such an ambitious title, my plan for
this talk is to locate these various developments within a
broader context.  How do they all fit together and relate to
each other?  What do they signify for the next decade of
particle physics?  And most important, what is the role of such
``low-energy'' high energy physics in the anticipated era of
new discoveries at the Tevatron and LHC?  In short, what do
we know now about the physics of flavor, and where do we hope
to go in the future?

\section{The Standard Model at low energies}

We begin by recalling what the Standard
Model looks like at ``low'' energies, by which I mean
renormalization scales $\mu$ below about 10 GeV.  I will
refer to this theory as the Low Energy Standard Model, or
LESM.  At these energies we have a theory with five quarks
($u,d,s,c,b$) and six leptons
($e,\mu,\tau,\nu_e,\nu_\mu,\nu_\tau$).  There is unbroken
$SU(3)_C\times U(1)_{EM}$ gauge symmetry, with eight massless
gluons and a massless photon.  There are Dirac masses for
the quarks and charged leptons, and perhaps also Majorana
masses for the neutrinos.  The gauge interactions and mass
terms are renormalizable operators in the effective
lagrangian.  With the exception of neutrino masses, these
interactions also conserve individual flavor quantum numbers.

In the LESM, all flavor changing processes are mediated by
{\it nonrenormalizable\/} interactions.  The most important
of these are four-fermion operators of mass dimension
six, of the form
\beq
{1\over M^2}\,\bar\psi_1\Gamma\psi_2\,
\bar\psi_3\Gamma\psi_4\,,
\eeqn
and fermion-gauge field interactions of dimension five, of the
form
\beq
{1\over M}\,\bar\psi_1\Gamma^{\mu\nu}\psi_2\,F_{\mu\nu}\,.
\eeqn
Here $\psi_i$ are fermion fields, $\Gamma$ and
$\Gamma^{\mu\nu}$ are combinations of Dirac matrices,
$F_{\mu\nu}$ is a color or electromagnetic gauge field
strength, and $M$ is a large mass scale.  For example,
neutron $\beta$ decay is mediated by the operator $\bar
u\gamma^\mu(1-\gamma^5)d\,\bar e\gamma_\mu(1-\gamma^5)\nu_e$,
the decay $K^-\to\pi^-\pi^0$ by $\bar
u\gamma^\mu(1-\gamma^5)s\,\bar d\gamma_\mu(1-\gamma^5)u$, and
$B^0-\bar B{}^0$ mixing by $\bar
b\gamma^\mu(1-\gamma^5)d\,\bar b\gamma_\mu(1-\gamma^5)d$. 
In each case, the scale $M$ is $M_W$.  There are
hundreds of such operators in the LESM,
responsible for an enormous variety of flavor changing
interactions.

In any effective field theory, nonrenormalizable operators are
generated by the exchange of heavy particles which have been
``integrated out'' of the theory at shorter distance scales. 
In the case of the LESM, some heavy particles which play this
role already have been observed directly at higher energies. 
First, we know that there is an electroweak gauge symmetry
which is realized nonlinearly,
$SU(2)_L\times U(1)_Y\to U(1)_{EM}$, with three group
generators broken by the the vacuum expectation value
$v=246\,$GeV.  These generators correspond to the
observed $W$ and $Z$ gauge bosons, with $M_W\simeq80\,$GeV and
$M_Z\simeq91\,$GeV.  Second, the $t$ quark has been seen, with
mass $m_t\simeq175\,$GeV.  Third, there are ``hard'' components
of the $SU(3)_C\times U(1)_{EM}$ gauge fields, with
$p^2>\mu^2$, which have also been removed from the theory and
their effects absorbed into the LESM lagrangian.  (Note that, 
except for the Goldstone bosons eaten by the $W$ and $Z$,
no Higgs particle or other remnant of the electroweak symmetry
breaking sector has been observed, and there is little
experimental evidence, even indirect, as to its nature.)

The couplings of the $W$, $Z$ and $t$ are fixed by gauge
symmetry, so virtual $W$, $Z$ and $t$ exchanges generate
nonrenormalizable operators in the LESM in a calculable way. 
Thus $\bar u\gamma^\mu(1-\gamma^5)d\,\bar
e\gamma_\mu(1-\gamma^5)\nu_e$, mediating neutron $\beta$
decay, is generated by $W$ exchange; $\bar
u\gamma^\mu(1-\gamma^5)s\,\bar d\gamma_\mu(1-\gamma^5)u$,
mediating $K^-\to\pi^-\pi^0$, is generated by $W$ exchange and
hard gluon loops; and $\bar
b\gamma^\mu(1-\gamma^5)d\,\bar b\gamma_\mu(1-\gamma^5)d$,
mediating $B^0-\bar B{}^0$ mixing, is generated by $W-t$ box
diagrams.

The interesting question, of course, is whether virtual $W$,
$Z$ and $t$ exchange can generate {\it all\/} flavor-changing
nonrenormalizable interactions in the LESM.  And if the
answer is no, which we hope and expect it to
be, then with what precision must we do experiments in order
to see deviations from this simple description?

There exist many models containing new particles and
interactions with masses in the range
$100\,{\rm GeV}<M<1\,$TeV.  A popular current framework for
such physics beyond the Standard Model is supersymmetry, which
in its minimal realization predicts a wealth of new degrees of
freedom in this region. There are squarks and sleptons,
gauginos, and charged and neutral Higgs bosons and their
superpartners.  The exchange of these heavy particles can have
the effect of changing the coefficients of operators in the
LESM which were already generated by $W$, $Z$ and $t$
exchange; for example, $\tilde t-\widetilde w$ box diagrams
generate a contribution to the operator $\bar
b\gamma^\mu(1-\gamma^5)d\,\bar b\gamma_\mu(1-\gamma^5)d$
responsible for $B^0-\bar B{}^0$ mixing.  They can also have
the effect of generating {\it new\/} nonrenormalizable
operators in the LESM; for example, charged Higgs exchange
generates the operator $\bar cb\,\bar\tau\nu_\tau$, which
would provide a scalar component to semileptonic $B$ decay. 
Although it is certainly not the only possibility, the
Minimal Supersymmetric Standard Model (MSSM) provides a
convenient paradigm for exploring the variety of new
interactions which new physics might induce at low
energies~\cite{GiudiceLP}.

\section{The quark sector of the Standard Model}

If we are to look in the LESM for a sign of new physics such as
the MSSM, the Standard Model contributions must be well
understood.  From here on, we will concentrate on flavor
changing processes in the quark sector, which has a
particularly rich phenomenology.  (With the recent strong
experimental hints for neutrino masses and mixing, the
phenomenology of the lepton sector is becoming interesting,
too~\cite{neutrinosLP}.)  There are three quark
$SU(2)$ doublets, which written in the mass eigenstate basis
are
\beq
\pmatrix{u\cr d},\qquad\pmatrix{c\cr s},\qquad
\pmatrix{t\cr b}.
\eeqn
The neutral current interactions of the quarks with the
$\gamma$ and $Z$ (and with the gluon) are flavor-diagonal by
the GIM mechanism.  The flavor-changing interactions are
mediated only by charged current interactions with the
$W^\pm$, of the form
\beq
\pmatrix{\bar u&\bar c&\bar t}\gamma^\mu(1-\gamma^5)
\vckm\pmatrix{d\cr s\cr b}W_\mu+\hc\,.
\eeqn
The $3\times3$
unitary matrix $\vckm$, due to Cabibbo, Kobayashi and
Maskawa~\cite{Cabibbo,Kobayashi:1973pm}, is
\beq
\vckm=\pmatrix{V_{ud}&V_{us}&V_{ub}\cr
V_{cd}&V_{cs}&V_{cb}\cr V_{td}&V_{ts}&V_{tb}\cr}\,.
\eeqn
The elements of $\vckm$ have a hierarchical structure, getting
smaller away from the diagonal: $V_{ud}$, $V_{cs}$ and
$V_{tb}$ are of order 1, $V_{us}$ and $V_{cd}$ are of order
$10^{-1}$, $V_{cb}$ and $V_{ts}$ are of order $10^{-2}$, and
$V_{ub}$ and $V_{td}$ are of order $10^{-3}$.  The parameters
of $\vckm$ originate in the couplings of the chiral quark
fields to the sector of the theory that breaks the
electroweak and global flavor symmetries and generates the
quark masses.  In the mimimal Standard Model, these are Yukawa
couplings to the Higgs field $\phi$,
\beq
\lambda^{ij}_u\,\bar Q_L^{\,i}\,\widetilde\phi\,u_R^j+
\lambda^{ij}_d\,\bar Q_L^{\,i}\,\phi\,d_R^j+\hc\,.
\eeqn
The 36 complex parameters $\lambda^{ij}_u$ and
$\lambda^{ij}_d$ may be reduced to 10 independent ones by
applying the $U(3)\times U(3)\times U(3)/U(1)_B$ global
symmetries of the quark kinetic terms.  In the mass basis,
these 10 quantities are the six quark masses
($m_u,m_d,m_s,m_c,m_b,m_t$), and four parameters which
characterize $\vckm$, including a complex phase.

A particularly convenient parameterization of $\vckm$, due
to Wolfenstein~\cite{Wolfenstein:1983yz}, is
\beq
\vckm=\pmatrix{1-\textstyle{1\over2}\lambda^2&\lambda
&A\lambda^3(\rho-{\rm i}\eta)\cr -\lambda&
1-\textstyle{1\over2}\lambda^2&A\lambda^2\cr
A\lambda^3(1-\rho-{\rm 
i}\eta)&-A\lambda^2&1}+\O(\lambda^4)\,.
\eeqn
The Wolfenstein parameterization exploits the hierarchy in
$\vckm$ by expanding explicitly in the small parameter
$\lambda=\sin\theta_C\simeq0.22$.  The other parameters, $A$,
$\rho$ and $\eta$, are expected to be of order one.  To this
order in $\lambda$, $\vckm$ is approximately unitary,
\beq
\vckm\vckm^\dagger=1+\O(\lambda^4)\,.
\eeqn
If necessary, one may keep higher order terms in the
expansion; when this is done, it is often useful to define
$\bar\rho=\rho(1-{1\over2}\lambda^2)$ and
$\bar\eta=\eta(1-{1\over2}\lambda^2)$.  

There is an unremovable complex phase in $\vckm$ when
$\eta\neq0$.  This phase induces complex couplings in the
charged current interactions, which leads to the possibility
that CP symmetry is violated.  Because CP violation is a
purely quantum phenomenon, it can be observed only in the
interference between different quark-level amplitudes which
mediate the same physical process.  From the point of view of
the LESM, there are two consequences.  First, the coefficients
of the nonrenormalizable flavor-changing operators must be
taken to be complex numbers.  Second, as a consequence of
these unremovable complex phases, CP violation is
something which is {\it naturally\/} present in fundamental
interactions.  {\it It is not a mystery.}  The only question
is whether it occurs at the level which one would expect in
the Standard Model.

The phases of the elements of $\vckm$ in the Wolfenstein
parameterization are
\beq
\pmatrix{1&1&{\rm e}^{-{\rm i}\gamma}
\cr 1&1&1\cr {\rm e}^{-{\rm i}\beta}&1&1}+\O(\lambda^2)\,,
\eeqn
which defines the angles $\beta$ and $\gamma$.   Note that
only the smallest elements $V_{ub}$ and
$V_{td}$ have phases of order one.  The other matrix elements
have small imaginary parts which appear only if $\vckm$ is
expanded to higher order in $\lambda$.  For example,
${\rm arg}\,V_{ts}=\lambda^2\eta$ is neglected at lowest
order.  As first pointed out by Kobayashi and
Maskawa~\cite{Kobayashi:1973pm}, and as is clear from this
parameterization, CP violation involves the interference
between amplitudes involving all three quark generations. 
Because of the smallness of $V_{ub}$ and
$V_{td}$, this phenomenon is inherently suppressed and
only observable in particular experimental situations.

The violation of CP symmetry was first observed in neutral
$K$ decays in 1964 by Cronin and Fitch.  More precisely, what
was observed was CP violation in $K^0-\bar K{}^0$ mixing,
inducing a small CP-even component of the $K_L$ and thereby
allowing the transition $K_L\to\pi\pi$.  Mixing in the $K$
system is mediated by a $\Delta S=2$ operator, $\bar
s\gamma^\mu(1-\gamma^5)d\,\bar s\gamma^\mu(1-\gamma^5)d$,
which is generated in the Standard Model by box diagrams
involving virtual $W$'s and $u$, $c$, and $t$ quarks.  The
coefficient of the $\Delta S=2$ operator is of the form
$AV_{cd}^2V_{cs}^{*2}+BV_{td}^2V_{ts}^{*2}
+CV_{cd}V_{td}V_{cs}^*V_{ts}^*$, for constants $A,B,C$. 
Although the terms proportional to $B$ and $C$ are
suppressed by powers of $\lambda$, they grow quickly with
$m_t$ and their contribution is important.  The
violation of CP depends on the fact that the overall
coefficient is complex, so the
$t$-mediated terms which bring in $V_{td}$ play the key
role.  The magnitude and phase of the phenomenological
parameter $\epsilon_K$ which characterizes CP violation in
mixing have been measured very precisely:
$|\epsilon_K|=[2.258\pm0.018]\times10^{-3}$ and
$\arg\epsilon_K=(43.49\pm0.08)^\circ$~\cite{Buchalla:1996vs}. 
However, predicting $\epsilon_K$ in the Standard Model is
problematic, because one needs not only the coefficient of the
$\Delta S=2$ operator but also its matrix element,
$\langle\bar K{}^0|\,\bar s\gamma^\mu(1-\gamma^5)d\,\bar
s\gamma^\mu(1-\gamma^5)d\,|K^0\rangle$.  Theoretical
calculations of this matrix element rely on lattice QCD, and
uncertainties are at the level of 20\%.  What can be said now
is only that $\epsilon_K$ is at the {\it right level\/} to
come from the phase in $\vckm$.  However, this is
still a significant observation!

More than three decades later, this year has seen the
definitive observation of CP violation in the $\Delta S=1$
sector of the LESM as well.  The operators in question are
generated in the Standard Model by strong and electroweak
penguin diagrams, the most important of which are
\beqa
&&Q_6=\bar s_\alpha\gamma^\mu(1-\gamma^5)d_\beta
\sum_{q=u,d,s}\bar q_\beta\gamma_\mu(1-\gamma^5)q_\alpha\,,\CR
&&Q_8=\bar s_\alpha\gamma^\mu(1-\gamma^5)d_\beta
\sum_{q=u,d,s}e_q\,\bar
q_\beta\gamma_\mu(1-\gamma^5)q_\alpha\,,
\eeqan
where $\alpha,\beta$ are color indices.  Since penguins
can include intermediate $t$ quarks, the operator coefficients
$C_6$ and $C_8$ receive complex contributions proportional to
${\rm Im}\,V_{td}V_{tb}^*$.  Both $Q_6$ and $Q_8$ mediate
the transition $K\to\pi\pi$, and because $C_6$ and $C_8$
are complex, they mediate the CP violating transition
$K_L\to\pi\pi$.  The phenomenological parameter $\epsilon'$
measures the extent to which CP violation differs in the
$\Delta I={1\over2}$ and $\Delta I={3\over2}$ channels, and a
nonzero $\epsilon'$ is a signal that CP violation in the
neutral $K$ system cannot be explained by mixing alone (in
which case it would be independent of the final state).

This year, both KTeV (${\rm
Re}[\epsilon'/\epsilon_K]=(28.0\pm4.1)
\times10^{-4}$~\cite{Alavi-Harati:1999xp})
and NA48 (${\rm
Re}[\epsilon'/\epsilon_K]=(18.5\pm7.3)
\times10^{-4}$~\cite{NA48epspreps})
confirmed the earlier NA31 result (${\rm
Re}[\epsilon'/\epsilon_K]=(23\pm7)
\times10^{-4}$~\cite{Barr:1993rx})
of a nonzero $\epsilon'/\epsilon_K$.  As with $\epsilon_K$,
however, the prediction of $\epsilon'/\epsilon_K$ in the
Standard Model is problematic due to hadronic uncertainties. 
The matrix elements
$\langle\pi\pi|\,Q_6\,|K_L\rangle$ and
$\langle\pi\pi|\,Q_6\,|K_L\rangle$ are not related to each
other, because of their different isopsin properties.  In a
very crude approximation which highlights the role of hadronic
uncertainties, one may write~\cite{Bosch:1999wr}
\beq
{\rm Re}[\epsilon'/\epsilon_K]\approx\Big[B_6^{1/2}
-0.5B_8^{3/2}\Big]\times10^{-3}\,,
\eeqn
where the $B_i$ are ``bag factors'' which parametrize the
matrix elements.  A guess based on the vacuum insertion ansatz
would yield $B_6^{1/2}=1.0$, $B_8^{3/2}=0.8$, and thus ${\rm
Re}[\epsilon'/\epsilon_K]\approx7\times10^{-4}$, smaller by a
factor of three than the experimental average.  This has led
some to speculate that the Standard Model cannot accommodate
such a large effect.  However, it cannot be stressed strongly
enough that the hadronic uncertainties are large and hard to
quantify, and that the values of $B_i$ taken from vacuum
insertion easily could be off by a factor of two or more in
either direction.  It is worth noting in this context that the
closely related hadronic matrix elements which account for the
$\Delta I={1\over2}$ rule in $K$ decays remain very poorly
understood.  Under the present circumstances, there is not
even a hint of new physics in this measurement.

In fact, what we have learned from the neutral $K$ system is
not that the Standard Model is ruled out, but that it is
confirmed: CP violation is at the level which one would
expect.  There is strong evidence for the correctness of
the Standard Model picture that CP violation arises from the
complex phase in $\vckm$, although hadronic uncertainties
preclude precision tests as yet.  In fact, the time has come to
adopt the point of view that CP violation is a natural part of
the Standard Model, and that as a phenomenon it is no longer
particularly interesting for its own sake.  Of course, there
do remain crucial questions.  Are there sources of CP
violation beyond those from $\vckm$, and how can we observe
them?  Can CP violation be a tool for exploring the breaking
of flavor symmetries in some more fundamental theory?  To
address these important issues, it is necessary to explore
the nonrenormalizable operators of the LESM in as much variety
and detail as possible.

\section{Constraints on $\vckm$}
\label{VCKMconstraints}

We already have quite a lot of experimental information on
the parameters of $\vckm$, interpreting the current
data within the framework of the Standard Model.  In
this section I review the current constraints, which have
changed little in the past year.

The best known element of $\vckm$ is
$V_{us}=\lambda=\sin\Theta_C$, the Cabibbo angle.  It is
proportional to the coefficient of the LESM operator $\bar
u\gamma^\mu(1-\gamma^5)s\,\bar e\gamma_\mu(1-\gamma^5)\nu_e$,
which mediates the well measured semileptonic decay
$K^-\to\pi^0e^-\bar\nu_e$.  To
relate the rate for this process to $\lambda$, one needs
to know the hadronic matrix element $\langle\pi^0|\,\bar
u\gamma^\mu(1-\gamma^5)s\,| K^-\rangle$.  Fortunately, in the
chiral $SU(3)$ limit this is the matrix element of a globally
conserved current, which can be computed exactly. 
Furthermore, chiral perturbation theory can be used to control
higher order corrections in $m_s$, so the hadronic physics is
under very good control~\cite{Donoghue:1992dd}.  From the
measured rate for  $K^-\to\pi^0e^-\bar\nu_e$, one
finds~\cite{Caso:1998tx}
\beq
\lambda=0.2196\pm0.0023\,,
\eeqn
an accuracy of about one percent.

The second best known element of $\vckm$ is
$V_{cb}=A\lambda^2$.  This constant is the coefficient of the
LESM operator $\bar c\gamma^\mu(1-\gamma^5)b\,\bar
e\gamma_\mu(1-\gamma^5)\nu_e$, responsible for semileptonic
$B$ meson decays.  Both the exclusive decay $\bar B\to
D^*\ell\,\bar\nu$ and the inclusive decay $\bar B\to
X_c\ell\,\bar\nu$ have been used to extract $V_{cb}$,
yielding consistent results.  In both cases, the Heavy Quark
Effective Theory, based on the approximation $m_b\to\infty$,
is used to control the hadronic physics~\cite{Ligeti99}.
Combining the various experimental determinations, one finds
\beq
V_{cb}=0.040\pm0.002\qquad\Rightarrow\qquad
A=0.83\pm0.04\,.
\eeqn
Theoretical uncertainties dominate the quoted error.  The
prospects for increasing the accuracy with which $A$ is known
are summarized nicely in Ref.~\cite{Ligeti99}.

\begin{figure}[htb]
\begin{center}
\vskip-1cm
\epsfig{file=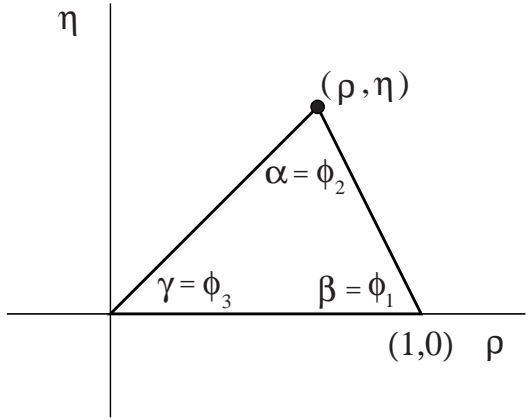,height=2.5in}
\caption{The unitarity triangle.}
\label{fig:ut}
\end{center}
\end{figure}

With $\lambda$ and $A$ reasonably well determined, $\rho$ and
$\eta$ remain the most important unknowns in $\vckm$.  It is
convenient to plot the point $\rho+{\rm i}\,\eta$ in the
complex plane.  The triangle which is made by connecting the
points $(0,0)$, $(\rho,\eta)$ and $(1,0)$ is known as the
``Unitarity Triangle'', because its closure may be related to
the unitarity of $\vckm$.  The angles of the Unitarity
Triangle are labeled as shown in Fig.~\ref{fig:ut}; on the
west side of the Pacific Ocean, they are known as
$(\phi_1,\phi_2,\phi_3)$, and on the east
side they are known respectively as
$(\beta,\alpha,\gamma)$.  The unitarity of $\vckm$ is often
expressed as the condition $\alpha+\beta+\gamma=\pi$.  When
$\vckm$ is expanded to higher order in $\lambda$, what is
plotted is often the point
$\bar\rho+{\rm i}\,\bar\eta$ rather than $\rho+{\rm i}\,\eta$.

There are three important constraints on $\rho$ and $\eta$. 
The first is from the measurement of $\epsilon_K$.  The
computation of the box diagram which generates the LESM
operator $\bar s\gamma^\mu(1-\gamma^5)d\,\bar
s\gamma^\mu(1-\gamma^5)d$ responsible for $K^0-\bar K{}^0$
mixing can include intermediate charm and top quarks, and so
the coefficient is a complicated combination of $\vckm$
parameters.  To lowest order in $\lambda$, $|\epsilon_K|$ may
be written as~\cite{Buras:1997fb}
\beq
|\epsilon_K|=3.8\times10^4\cdot B_K\cdot A^2\lambda^6
\cdot\eta\,
\Big[f(x_c,x_t)+A^2\lambda^4(1-\rho)g(x_t)\Big]\,,
\eeq{epsk}
where the short-distance quantities
$f(x_c,x_t)\approx7.2\times10^{-4}$ and
$g(x_t)\approx1.35$ are functions of
$x_{c,t}=m_{c,t}^2/M_W^2$.  They have been computed to next-to-leading order
in QCD.  The two terms in Eq.~\leqn{epsk} are of roughly the
same magnitude because of the strong dependence of the box
diagram on $x_t$.  The result is a hyperbola in the
$(\rho,\eta)$ plane. However, the bag factor $B_K$, which
parameterizes the hadronic matrix element $\langle\bar
K{}^0|\,\bar s\gamma^\mu(1-\gamma^5)d\,\bar
s\gamma_\mu(1-\gamma^5)d\,|K^0\rangle={8\over3}m^2_Kf_K^2B_K$,
is not known precisely.  Lattice QCD estimates yield
approximately the range $0.6<B_K<1.0$, but the dominant
uncertainties are from quenching and have not been quantified
reliably~\cite{Sharpe:1998hh}.  Also, note that the fourth
power of $A$ appears in the second term of Eq.~\leqn{epsk}, so
the 5\% experimental uncertainty on $A$ is magnified
considerably.

The second constraint comes from the measurement of
$|V_{ub}/V_{cb}|=\lambda\sqrt{\rho^2+\eta^2}$.  The parameter
$V_{ub}$ is most easily extracted from semileptonic $B$
decays, since it is proportional to the coefficient of the
LESM operator $\bar u\gamma^\mu(1-\gamma^5)b\,
\bar\ell\gamma_\mu(1-\gamma^5)\nu_\ell$.  An important
experimental difficulty is that there is a huge background
from the process $b\to c\ell\,\bar\nu_\ell$, which in the
Standard Model is approximately 100 times larger.  Rejecting
this background requires one to restrict the analysis either
to a specific hadronic final state or to a small fraction of
the lepton phase space.  As
will be discussed in Section~\ref{Vubsection}, either
approach introduces significant model dependence into the
hadronic physics, with uncertainties which are very hard to
quantify meaningfully.  Taking the current central values but
with a reasonably conservative attitude toward the theoretical
errors, one has
$|V_{ub}/V_{cb}|=0.090\pm0.025$~\cite{Ligeti99}.  Analyses
with subtantially smaller theoretical uncertainties should be
taken with a grain of salt.

The third important constraint comes from $B^0-\bar B{}^0$
mixing, which is mediated by the LESM operator 
$\bar b\gamma^\mu(1-\gamma^5)d\,
\bar b\gamma^\mu(1-\gamma^5)d$.
In the Standard Model, this operator is generated by $t-W$
box diagrams, with a coefficient proportional to
$|V_{td}{}^*V_{tb}|^2$.  The
phenomenological parameter $\Delta m_d$ is precisely
measured, $\Delta m_d=0.464\pm0.018\,{\rm
ps}^{-1}$~\cite{Caso:1998tx}.  However, as in the case of
$K^0-\bar K{}^0$ mixing, relating this number to fundamental
quantities requires hadronic matrix elements which are
difficult to compute.  At leading order in $\lambda$ and
next-to-leading order in QCD, one finds~\cite{Buras:1997fb}
\beq
\Delta m_d=1.30\,G_F^2M_W^2/6\pi^2\cdot m_{B_d}
\cdot  f_{B_d}^2B_{B_d}\cdot
A^2\lambda^6\cdot[(1-\rho)^2+\eta^2]\,.
\eeqn
The quantity $f_{B_d}^2B_{B_d}$ parameterizes the matrix
element.  It is most accurately computed on the lattice,  where
$f_{B_d}\sqrt{B_{B_d}}$ has an uncertainty at the level of
20\%~\cite{Sharpe:1998hh}.

In this case, the hadronic uncertainties can be reduced if
the analagous quantity $\Delta m_s$ can be measured in
$B_s-\bar B_s$ mixing.  Then one can form a ratio,
\beq
{\Delta m_d\over\Delta
m_s}=\xi^{-2}\,{|V_{td}|^2\over|V_{ts}|^2}
=\xi^{-2}\cdot\lambda^2\,[(1-\rho)^2+\eta^2]\,,
\eeqn
where $\xi=[f^2_{B_s}B_{B_s}/f^2_{B_d}B_{B_d}]^{1/2}$.
The ratio $\xi$ is unity in the $SU(3)$ limit. It can be
studied much more reliably on the lattice than can either
the numerator or the denominator, although calculations are
still done only in the quenched theory.  Recent estimates
which attempt to include quenching errors give
$\xi=1.14\pm0.13$~\cite{Sharpe:1998hh}.  The current limit on
$B_s-\bar B_s$ mixing, $\Delta m_s>12.4\,{\rm
ps}^{-1}$~\cite{Buchalla:1996vs}, already contributes to
constraining $(\rho,\eta)$.  As the lower bound on $\Delta m_s$
increases (preliminary reports at this conference indicate
$\Delta m_s>14.3\,{\rm ps}^{-1}$ at 95\%
C.L.~\cite{BlaylockLP}), this constraint will become ever
more important.

\begin{figure}[htb]
\begin{center}
\vskip-0.5cm
\epsfig{file=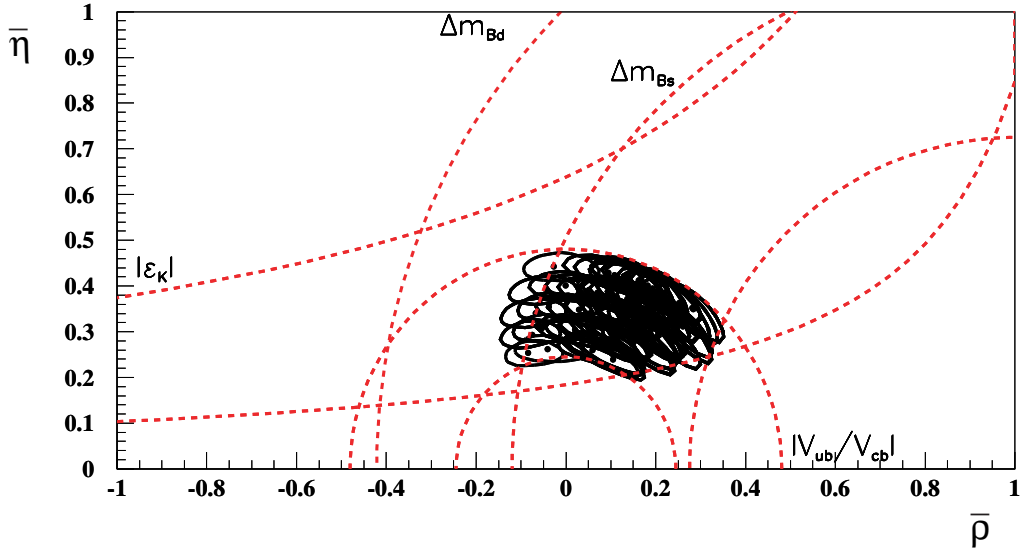,width=6in}
\vskip-7.25cm
\caption{Constraints on the unitarity triangle, taken from
{\sl The Babar Physics Book}, 1998~\cite{Harrison:1998yr}. 
The constraints are presented in the $(\bar\rho,\bar\eta)$
plane, in which subleading corrections of relative order
$\lambda^2$ have been included.  Theoretical uncertainties are
treated with a scanning method.}
\label{fig:constraints}
\end{center}
\end{figure}

Combining the limits in the $(\rho,\eta)$ plane from
$\epsilon_K$, $|V_{ub}|$, $\Delta m_d$ and $\Delta m_s$
requires a consistent treatment of experimental and
theoretical errors.  This is problematic because the
dominant theoretical hadronic uncertainties are difficult to
quantify meaningfully and are certainly not in any sense
Gaussian distributed about their ``central values''.  The
BaBar collaboration has proposed that a scanning method
be used to deal with the theoretical errors associated with
$B_K$, $f_{B_d}^2B_{B_d}$, $\xi$, and the extraction of
$V_{ub}$.  In this approach, the global fit to the data is
repeated for a range of values of the theoretical inputs, and
then the various allowed regions are overlayed to obtain an
overall constraint.  {\sl The BaBar Physics Book} uses the
fairly conservative ranges $B_K=0.6-1.0$,
$f_{B_d}^2B_{B_d}=160-240\,$MeV, $\xi=1.06-1.22$, and
$|V_{ub}/V_{cb}|=0.06 - 0.10$~\cite{Harrison:1998yr}.  Their
allowed region for $(\rho,\eta)$, at 95\% C.L., is reproduced
in Fig.~\ref{fig:constraints}. 

Other, more restrictive, regions in the $(\rho,\eta)$ plane
have appeared in the literature~\cite{Parodi:1999nr}. 
However, it is important to note that they differ from the one
shown here almost entirely in a less conservative treatment of
theoretical uncertainties.  With the exception of new
determinations of $V_{ub}$, which are consistent with previous
measurements and with comparable errors, there has been no
recent change in the {\it experimental\/} input into the
constraints.  Until the lattice calculations of hadronic
matrix elements improve, with quenching corrections under
serious control, a conservative treatment of theoretical
uncertainties is appropriate.  The prospects for
reducing the model dependence in the extraction of $V_{ub}$
will be discussed below; here both theory and experiment
could contribute to improvement in the situation.  Until
there are future developments,
Fig.~\ref{fig:constraints} remains a reasonable
representation of the state of our knowledge of $\vckm$.

\section{Electroweak baryogenesis}

An interesting motivation to study CP violation
in the Standard Model is the possible connection to the
generation of the cosmological asymmetry between matter and
antimatter.  The net baryon-to-photon ratio in the universe is
small but almost certainly nonzero, $(n_B-n_{\bar
B})/n_\gamma\simeq 3\times10^{-10}$.  While it is possible that
this asymmetry is an initial condition of the Big Bang, it
would be more satisfying if it could be explained as
arising dynamically from matter symmetric initial conditions. 
It would be even more satisfying if the dynamics could be
understood in terms of physical processes which already are
known to occur in the early universe.

Sakharov identified three conditions which must be
satisfied simultaneously for a baryon number asymmetry to be
generated dynamically~\cite{Sakharov67}.  First, at
some point the universe must depart from a state of thermal
equilibrium.  Second, at this time there must be unsuppressed
processes which violate baryon number.  Third, there must be
C and CP violation in the theory, to allow baryons and
antibaryons to be created at different rates.  Finally, given
these ingredients there also must exist a mechanism to
generate the asymmetry at the right level.

It turns out that the Standard Model itself satisfies
Sakharov's conditions during the electroweak phase
transition~\cite{Trodden:1998ym}.  At temperatures
$k_BT\simeq100\,$GeV, the Higgs sector settles into its ground
state and assumes a vacuum expectation value
$\langle\phi\rangle=v=246\,$GeV, breaking the electroweak
gauge symmetry $SU(2)_L\times U(1)_Y\to U(1)_{EM}$.  If the
phase transition is first order, it will proceed by the
nucleation of bubbles of ``true'' vacuum,
$\langle\phi\rangle=v$, in the ``false'' vacuum background
with $\langle\phi\rangle=0$.  The walls of the expanding
bubbles will be sites of local thermal nonequilibrium, at which
baryogenesis could take place.  The baryon number violating
interactions are provided by sphalerons, classical
configurations of nonzero winding of the gauge and Higgs
fields.  Sphalerons are unsuppressed at high temperatures,
when the electroweak symmetry is unbroken; at lower
temperatures, they are suppressed by $\exp(-v/k_BT)$ and
rapidly become unimportant as the universe cools.  Finally,
C is violated maximally in the Standard Model, while CP
violation is supplied by $\vckm$.

Whether this mechanism is sufficient to produce the observed
baryon number asymmetry is a detailed question which depends
on the dynamics occurring at the bubble wall.  In one
scenario, the baryons are generated by a difference in
scattering of left- and right-handed top quarks from the
moving wall.  The chiral asymmetry is then converted by
sphalerons in the false vacuum into a net baryon number.  Here
it is crucial that there be enough CP violation in the
quark-Higgs interactions. In addition, it is necessary that
the asymmetry not be subsequently washed out by insufficiently
suppressed sphaleron-mediated interactions in the bubble wall
itself.  This latter condition requires that the phase
transition be strongly first order.

It turns out that the minimal Standard Model fails on both
counts.  First, because the effects of CP violation are
suppressed both by the smallest couplings in $\vckm$ and by
the quark masses, the ``natural'' size of the baryon
asymmetry which could be generated is
$10^{-21}$~\cite{Trodden:1998ym}.  To produce the observed
asymmetry would require an enhancement of approximately eight
orders of magnitude.  Second, it is now known that the
electroweak phase transition is not sufficiently first order
to prevent sphalerons from eradicating the baryon number
asymmetry once it is generated~\cite{Trodden:1998ym}.  New
physics is needed, if the asymmetry is to be produced at the
electroweak phase transition.

Perhaps the most natural candidate for this new physics is
supersymmetry.  It is possible in the MSSM for the $\tilde
t$ squark and the neutral and charged scalars $h$ and $H^\pm$
to be light, with masses in the range
$100-200\,$GeV.  In particular, if $100\,{\rm GeV}<m_{\tilde
t_R}<m_t$ and $m_h<115\,$GeV, then the phase transition can be
sufficiently first order.  There are also many new sources of
CP violation in the MSSM.  The most important of these for
electroweak baryogenesis is the phase in the $H^\pm$ mass
matrix.  If this phase is of order one, and the $\tilde t$ and
$H^\pm$ are light enough, then it remains possible to generate
the observed baryon asymmetry~\cite{Trodden:1998ym}.

However, even a large CP violating phase in the $H^\pm$
mass matrix will have small CP violating effects in $B$
decays.  This is because the new phase is in the $\Delta B=0$
sector of the lagrangian.  In such a scenario, it could be
that the only effect in the $B$ system is a large
contribution to $B^0-\bar B{}^0$ mixing from $\tilde
t-\widetilde w$ box diagrams.  This contribution could be as
large as the $t-W$ box diagrams of the Standard Model; in this
case, not only would the magnitude of $\Delta m_d$ be
affected, but also its phase would change from its Standard
Model value of $2\beta$.  Either of these effects, if large
enough, could be identified experimentally.

In summary, then, electroweak baryogenesis should be
considered as an indirect motivation for studying CP
violation in the $B$ system.  There are scenarios of
electroweak baryogenesis which would leave a signature in
$B^0-\bar B{}^0$ mixing, but there is no reason to expect that
the new sources of CP violation which are needed to make these
scnearios viable would manifest themselves directly in $B$
decays.

\section{Future scientific program}

The question to be addressed in the next round of experiments
is whether the Standard Model can account completely for the
flavor changing operators of the LESM.  The hope, of course,
is that the answer to this question is no, and that in seeing
deviations from the Standard Model predictions we will get the
first hints of new physics at the TeV scale.

What will be required to address this issue will be a robust
program of $K$ and $B$ physics over the next decade.  In
particular, it would be ideal to measure $(\rho,\eta)$
independently in the $K$ and $B$ systems, through,
respectively, the coefficients of $\Delta S=1,2$ and $\Delta
B=1,2$ operators in the LESM.  It will especially important
that constraints on $\rho$ and $\eta$ be theoretically clean,
in the sense that they be free of model-dependence and
uncontrolled uncertainties associated with hadronic physics. 
Measurements which satisfy this criterion will have a crucial
role to play in uncovering physics beyond the Standard Model.

\subsection{$K$ physics}

In the $K$ system, there are two experiments of
particularly importance for constraining
$\vckm$.  The first is the measurement of the rate for
the process $K^-\to\pi^-\nu\bar\nu$.  This decay is
mediated by the $\Delta S=1$ operator $\bar
d\gamma^\mu(1-\gamma^5)s\,\bar\nu\gamma_\mu(1-\gamma^5)\nu$,
which in the Standard Model is generated dominantly by penguin
and box diagrams with virtual $t$ and $c$ quarks.  The
amplitude depends on a linear combination of
$V_{td}{}^*V_{ts}$ and $V_{cd}{}^*V_{cs}$.  Because this is
a semileptonic decay, the hadronic matrix element
$\langle\pi^-|\,\bar d\gamma^\mu(1-\gamma^5)s\,|K^-\rangle$
can be related to the quantity $\langle\pi^0|\,\bar
u\gamma^\mu(1-\gamma^5)s\,| K^-\rangle$ measured in
$K^-\to\pi^0e^-\bar\nu$ and is under good theoretical control. 
The largest uncertainty arises from computing the
relative contribution from the virtual $c$ quark, and
this is primarily because of the experimental
uncertainty in $A$.  An analysis at leading order in
$\lambda$ and next-to-leading order in QCD yields the
expression~\cite{Buras:1997fb}
\beq
\BR(K^-\to\pi^-\nu\bar\nu)=3.4\times10^{-4}\cdot A^4
\lambda^{10}\cdot\left[\eta^2+(1+\delta_c-\rho)^2\right]\,.
\eeqn
Hence a measurement of the rate would constrain an annulus
centered at $(\rho,\eta)=(1+\delta_c,0)$  The charm
contribution is
\beq
\delta_c=1.5\times10^{-6}\cdot A^{-2}\lambda^{-4}\simeq0.40\,.
\eeqn
It has a theoretical error of approximately 15\%, which is
dominated by the remaining uncertainties in $A$ and, to a
lesser extent, $\lambda$.  Improvements in these quantities
would also affect the normalization of the overall result.

The second key experiment would be to measure the rate for
$K^0\to\pi^0\nu\bar\nu$.  Except for a small component of
higher angular momentum, this decay is purely CP
violating, with an amplitude proportional to ${\rm
Im}\,V_{td}{}^*V_{ts}=A^2\lambda^5\eta$.  There is no
contribution from intermediate charm quarks, and the
calculation is theoretically very clean.  At next-to-leading
order in QCD, one has~\cite{Buras:1997fb}
\beq
\BR(K_L\to\pi^0\nu\bar\nu)=1.5\times10^{-3}\cdot A^4
\lambda^{10}\cdot\eta^2\,.
\eeqn
A measurement of the rate would constrain a horizontal band in
the $(\rho,\eta)$ plane.

The only $\Delta S=2$ observable in the $K$ system is
$\epsilon_K$; the difficulties in using it to constrain
$\vckm$ have been discussed in Section~\ref{VCKMconstraints}. 
The theoretical expression for $\epsilon_K$ is given by
Eq.~\leqn{epsk}.  A reliable lattice computation of the bag
parameter $B_K$ is desperately needed, but this will have to
wait until the quenching corrections are understood
quantitatively and reliably.  A better determination of
$A$ is also very important.

\begin{figure}[htb]
\begin{center}
\vskip-0.8in
\hskip-1.5in
\epsfig{file=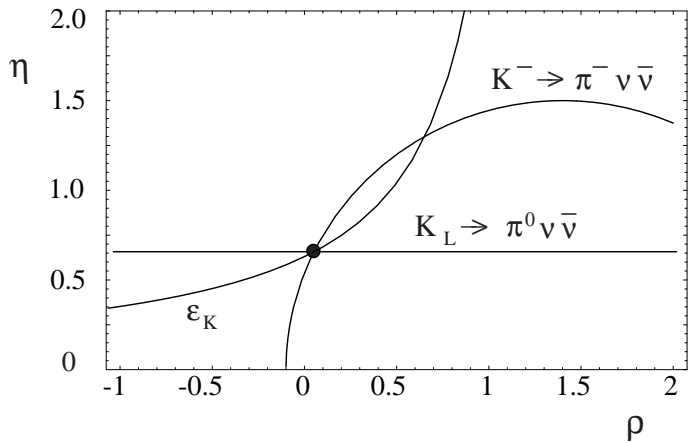,height=3in}
\caption{Hypothetical constraints on $(\rho,\eta)$
from measurements in the $K$ system.}
\label{fig:Kconstraints}
\end{center}
\end{figure}

The ability of $\BR(K^-\to\pi^-\nu\bar\nu)$,
$\BR(K_L\to\pi^0\nu\bar\nu)$ and $\epsilon_K$ to pinpoint
$(\rho,\eta)$ using $K$ physics alone is illustrated in
Fig.~\ref{fig:Kconstraints}.  This is an idealized
situation in which theoretical and experimental errors are
neglected, but nevertheless it shows the power of the 
experiments on rare $K$ decays to constrain the Standard Model.
As is apparent from the figure, of particular importance is the
process $\BR(K_L\to\pi^0\nu\bar\nu)$.  It should also be noted
that the error in normalization from the remaining uncertainty
in $A$ is largely correlated in the three quantities.

What are the experimental prospects for carrying out this
program?  In the mode $K^\pm\to\pi^\pm\nu\bar\nu$, the
program is underway, as one event already has been observed at
the Brookhaven expriment E787~\cite{Adler:1997am}.  The
observation of this single event is consistent with the
Standard Model expectation for the branching ratio at the
level of $10^{-10}$.  There is a proposal for a follow-on
experiment at Brookhaven, E949, with the goal of seeing
$10-20$ events.  A next generation experiment, CKM, has been
proposed to run at a future Fermilab Main Injector fixed
target program.  If successful, CKM will make a $10\%$
measurement of the branching ratio.

The mode $K_L\to\pi^0\nu\bar\nu$ is harder, both because the
branching ratio is expected to be an order of magnitude
smaller and because there are no charged particles in the
final state.  Proposals to attempt this measurement are being
developed at Brookhaven (BNL--E926) and Fermilab (KAMI).  The
two proposals take different approaches to achieving the high
level of background rejection which will be necessary for an
experiment to be feasible.  It is still unclear which
approach, if either, has the better chance to succeed.

\subsection{$B$ physics}

The $B$ system offers many opportunities to constrain the
Unitarity Triangle.  There are two classes of measurements:
those that measure the magnitudes of the elements of $\vckm$,
such as $V_{ub}$, $V_{cb}$, $\Delta m_d$ and $\Delta m_s$, and
those that measure directly the angles $\alpha$, $\beta$, and
$\gamma$.  The latter experiments probe CP violation in the
$B$ system directly.  Among their advantages is that in
certain cases they can be free of uncertainties from hadronic
physics.  Rather than provide a broad survey of $B$ decay
modes, which can be found in many places
elsewhere (see, for example, Ref.~\cite{Harrison:1998yr}), I
will focus on a few particularly important examples.  In so
doing , I will highlight what is meant by the measurement of a
rate or an asymmetry being ``theoretically clean''.

The immediate future for $B$ physics experiments is
extraordinarily exciting.  This year saw the commissioning of
two new asymmetric $B$ Factories operating at the
$\Upsilon(4S)$ resonance, PEP--II at SLAC and KEK--B at KEK. 
The luminosity goal for these accelerators, along with that
of the upgraded symmetric collider CESR at Cornell, is in the
range $10^{33}-10^{34}\,{\rm cm}^{-2}{\rm sec}^{-1}$.  The
corresponding detectors, BaBar, BELLE and CLEO--III, will
collect tens of millions of reconstructed $B$ decays.  In
addition, Run II at the Fermilab Main Injector will see
luminosities of the order of $10^{33}\,{\rm cm}^{-2}{\rm
sec}^{-1}$ in $p\bar p$ collisions at $2\,$TeV, and with the
substantial upgrades to CDF and D0, the hadron collider
detectors will make important contributions as well.  In
particular, CDF expects to measure $\Delta m_s$ with high
precision over and well beyond its entire Standard Model range.

In the longer term, there are proposals for dedicated $B$
experiments at the Tevatron (BTeV) and the LHC (LHC--B). 
Such detectors, installed in the far forward region, would
see enormous fluxes of all $b$ flavored hadrons.  How best to
do $B$ physics in such an active environment is currently the
subject of intense study.

\subsubsection{CP asymmetry in $B\to J/\psi\, K_S$}

The ``gold-plated'' measurement in the $B$ system is the CP
asymmetry in $B\to J/\psi K_S$.  Not only is the final state
easy to identify and the branching ratio relatively large
($\sim 5\times10^{-4}$), but the calculation of the asymmetry
is theoretically quite clean.  This measurement will be one
of the early goals of all the $B$ Factory experiments.

The CP asymmetry arises from the interference between the
direct decay $B^0\to J/\psi\, K_S$ and the mixing-induced decay
$B^0\to\bar B{}^0\to J/\psi\, K_S$.  One can measure either a
time-dependent or a time-integrated CP asymmetry.  In the
first case, one measures
\beqa
a_{CP}(t)&=&{\Gamma(B_{\rm phys}^0(t)\to\psi K_S)-
\Gamma(\bar B{}^0_{\rm phys}(t)\to\psi K_S)\over
\Gamma(B_{\rm phys}^0(t)\to\psi K_S)+
\Gamma(\bar B{}^0_{\rm phys}(t)\to\psi K_S)}\CR
&=&{(1-|\lambda_{CP}|^2)\cos(\Delta m_d\, t)-
2{\rm Im}\,\lambda_{CP}\,\sin(\Delta m_d\, t)\over 
1+|\lambda_{CP}|^2}\CR
&=&-{\rm Im}\,\lambda_{CP}\,\sin(\Delta m_d\, t)\,,
\eeqa{asymtime}
where $\lambda_{CP}$ is a CP violating observable which
depends on the final state, and the last line holds because in
the case of $J/\psi\,K_S$ direct CP violation is negligible
and therefore $|\lambda_{CP}|=1$.  The notation $B^0_{\rm
phys}(t)$ refers to a state which is tagged as a $B^0$ at time
$t=0$ and decays at time $t$.  For the time-integrated
asymmetry (which cannot be observed at the
$\Upsilon(4S)$ because of the quantum coherence of the initial
$B^0\bar B{}^0$ state), one has , with $|\lambda_{CP}|=1$,
\beq
a_{CP}={\Gamma(B^0\to\psi K_S)-
\Gamma(\bar B{}^0\to\psi K_S)\over
\Gamma(B^0\to\psi K_S)+\Gamma(\bar B{}^0\to\psi K_S)}=
-{x_d\over1+x_d^2}\,{\rm Im}\,\lambda_{CP}\,,
\eeq{asymint}
where $x_d=\Delta m_d/\Gamma(B_d)=0.72\pm0.03$.

The phase of $\lambda_{CP}$ is independent of the phase
convention for $\vckm$ and has a straightforward relationship
to the underlying process. Let $\M_B$ be the amplitude for
$B^0-\bar B{}^0$ mixing, $\A_{\psi K_S}$ the amplitude
for $B^0\to J/\psi\, K_S$, and $\M_K$ the amplitude for
$K^0-\bar K{}^0$ mixing.  Then
\beq
\arg\lambda_{CP}=\arg\M_B+2\arg\A_{\psi K_S}+\arg\M_K+\pi\,,
\eeq{lampsiks}
where the extra $\pi$ arises because the final state
$J/\psi\, K_S$ is CP-odd.  In the Standard Model, $\M_B$ is
dominated by $t-W$ box diagrams,
and $\arg\M_B=2\arg(V_{td}V^*_{tb})=-2\beta$.
Similarly, $\M_K$ is mediated by $c-W$ box diagrams, so
$\arg\M_K=2\arg(V_{cd}V^*_{cs})=0$.
Finally, the dominant contribution to the decay amplitude is
from the LESM operator $\bar c\gamma^\mu(1-\gamma^5)b\,\bar
s\gamma_\mu(1-\gamma^5)c$, induced by tree level $W$ exchange,
so $\arg\A_{\psi K_S}=\arg(V_{cb}V^*_{cs})=0$.
Putting these elements together, we find
$a_{CP}\propto{\rm Im}\,\lambda_{CP}=\sin2\beta$.  Note that
as long as the arguments of $\M_K$ and $\A_{\psi K_S}$ vanish
as in the Standard Model, the asymmetry in $B\to
J/\psi\,K_S$ measures the phase of $B^0-\bar B{}^0$ mixing.

The tree process is not the only one which can contribute to
$\A_{\psi K_S}$.  There are also strong penguin diagrams,
which induce in the LESM the operator $\bar s\gamma^\mu
T^ab\,\bar c\gamma_\mu T^ac$, where $T^a$ are $SU(3)$
matrices.  The matrix elements $\langle\psi K_S|\,\bar
c\gamma^\mu(1-\gamma^5)b\,\bar s\gamma_\mu(1-\gamma^5)c\,
|B\rangle$ and $\langle\psi K_S|\,\bar s\gamma^\mu
T^ab\,\bar c\gamma_\mu T^ac\,|B\rangle$ are independent
hadronic quantities which are not related to each other by
any symmetry.  Therefore the ratio of the penguin-mediated
amplitude to the tree amplitude is model-dependent and
cannot as yet be estimated reliably.  This is potentially a
source of serious trouble, because it is absolutely
indispensable to know accurately the weak phase of $\A_{\psi
K_S}$, which depends on this ratio.

However, let us examine the penguin contribution more closely.
The largest diagram has a virtual $t$ quark and is
proportional to $V_{tb}V_{ts}^*$.  Fortunately, though,
$\arg(V_{tb}V^*_{ts})=\O(\lambda^2)$ is almost the same as
$\arg(V_{cb}V^*_{cs})$.  Since the penguin diagram is also
loop suppressed by $\alpha_s/4\pi$, the phase of
$\A_{\psi K_S}$ is largely unaffected by this term, {\it
regardless of the unknown ratio of hadronic matrix
elements.}  The penguin diagram with a virtual $u$ quark has
an effect of the same small size; its argument $-\gamma$ is of
$\O(1)$, but its magnitude is suppressed by $\lambda^2$
compared to the $t$ penguin.  In the end, penguin corrections
to $\arg\A_{\psi K_S}$ are expected to be below the level of
$10^{-2}$ and can be safely neglected.  This is what is meant
by the statement that $B\to J/\psi\,K_S$ is a ``theoretically
clean'' mode from which to extract a CP asymmetry.

The current limits on $(\rho,\eta)$ place significant
constraints on the values of $\sin2\beta$ which are
consistent with the Standard Model.  Treating theoretical
uncertainties conservatively, one finds an allowed range of
roughly $0.4<\sin2\beta<0.8$~\cite{Harrison:1998yr}.  Note
that the sign of the asymmetry is predicted. 
The recent CDF result is
$$
\sin2\beta=0.79^{+0.41}_{-0.44}\,,\qquad{\rm with}\qquad
0.00\le\sin2\beta<1.00\quad{\rm at}\ 93\%\ {\rm C.L.}\,.
$$
Although the errors are still large, the measurement is
consistent with the indirect constraints, and most important,
the sign comes out as expected.   This result is actually a
little disappointing, since if the sign had been negative then
we would have had our first unambiguous indication of physics
beyond the Standard Model.

Future measurements will constrain the asymmetry in $B\to
J/\psi\,K_S$ much more tightly.  The upcoming $B$ Factory
experiments BaBar and BELLE, along with CDF and D0 at Run II
and HERA--B, will succeed in achieving
$\delta\sin2\beta=0.05-0.10$.  Farther in the future, hadronic
``$B$ Factories'' should be able to identify this final
state with little trouble and measure the asymmetry extremely
well.  BTeV and LHC--B, as well as ATLAS and CMS, expect an
eventual accuracy of $\delta\sin2\beta=0.01-0.02$.  With
the theoretical prediction well under control and the
experimental prospects so promising, this mode is
``gold-plated'', indeed!

\subsubsection{CP asymmetry in $B\to\pi^+\pi^-$}

Unfortunately, the same cannot be said of $B\to\pi^+\pi^-$, a
mode once thought to be as useful as $B\to J/\psi\,K_S$ for
constraining the Unitarity Triangle.  An asymmetry analogous to
\leqn{asymtime} or
\leqn{asymint} may be measured for the $\pi^+\pi^-$ final
state.  The only difference in the analysis is that now
\beq
\arg\lambda_{CP}=\arg\M_B+2\arg\A_{\pi\pi}\,.
\eeq{lampipi}
The tree level contribution to $\A_{\pi\pi}$ is proportional
to $V_{ub}V^*_{ud}$.  Since $\arg(V_{ub}V^*_{ud})=-\gamma$,
the asymmetry in $B\to\pi^+\pi^-$ is proportional to ${\rm
Im}\,\lambda_{CP}=-\sin(2\beta+2\gamma)=\sin2\alpha$, using
the unitarity relation $\alpha+\beta+\gamma=\pi$.

The problem comes when we consider the penguin contributions
to $\arg\A_{\pi\pi}$.  The penguin diagram with an virtual $t$
quark in the loop yields a term in the amplitude with
argument $\arg(V_{tb}V^*_{td})=\beta$ rather than
$\arg(V_{ub}V^*_{ud})=-\gamma$, but with no suppression
by powers of $\lambda$.  The tree operator $\bar
u\gamma^\mu(1-\gamma^5)b\,\bar d\gamma_\mu(1-\gamma^5)u$ and
the penguin operator $\bar d\gamma^\mu T^ab\,\bar u\gamma_\mu
T^au$ have different weak phases and hadronic matrix
elements.  The asymmetry in $B\to\pi^+\pi^-$ still depends
cleanly on the weak phase of $\A_{\pi\pi}$, but $\A_{\pi\pi}$
is now proportional to an unknown linear combination of ${\rm
e}^{-\rm i\gamma}$ and ${\rm e}^{\rm i\beta}$.  Absent reliable
knowledge of both the relative {\it magnitudes\/} and {\it
strong phases\/} of
$\langle\pi^+\pi^-|\,\bar u\gamma^\mu(1-\gamma^5)b\,\bar
d\gamma_\mu(1-\gamma^5)u\,|B
\rangle$ and $\langle\pi^+\pi^-|\,\bar d\gamma^\mu T^ab\,\bar
u\gamma_\mu T^au\,|B\rangle$, even a precise measurement of
the asymmetry cannot be used to extract cleanly the phase of
any single operator in the LESM, nor the phase of any element
of $\vckm$.

Only the loop factor $\alpha_s/4\pi$ remains to
suppress the contribution of the penguin diagram.  Since the
decay $B\to K\pi$ is probably dominated by penguins
(in this mode the tree contribution to $b\to u\bar us$ is
suppressed by $\lambda^2$ compared to the penguin
contribution), it provides a probe of the strength of the
penguin matrix elements.  Recent data on $B\to K\pi$ which
indicate that $\BR(B\to K\pi)>\BR(B\to\pi\pi)$ are not
encouraging.  Denoting by $|P/T|$ the relative contribution of
penguin to tree processes in $B\to\pi\pi$, the data suggest
$0.2<|P/T|<0.6$~\cite{BlaylockLP,Harrison:1998yr}.  This is
well in the range where the ``penguin pollution'' is fatal.

In principle, the penguin contributions can be disentangled
by exploiting the isospin structure of the decay.  The tree
operator has both $\Delta I={1\over2}$ and $\Delta I={3\over2}$
components, while the penguin is pure $\Delta I={1\over2}$. 
Gronau and London observed that this could be done by
measuring the flavor-tagged rates for
$B\to\pi^+\pi^-,\pi^0\pi^0,\pi^\pm\pi^0$~\cite{Gronau:1990ka}. 
The experimental difficulty is with $B\to\pi^0\pi^0$:  not only
is the branching ratio for this ``color-suppressed'' mode
expected to be as small as $10^{-7}$, but the
four-photon final state is extremely difficult to identify. 
So far, none of the current or proposed $B$ Factories has
claimed to be able to make this measurement.

An alternative explored by Quinn and Snyder is to do a
Dalitz plot analysis of the CP asymmetry in
$B\to(\rho^-\pi^+,\rho^+\pi^-,\rho^0\pi^0)
\to\pi^+\pi^-\pi^0$~\cite{Snyder:1993mx}. The various
$\rho\pi$ intermediate states have different isospin quantum
numbers and hence different sensitivity to the tree and
penguin operators.  An additional advantage of this approach
is that one simultaneously would determine $\sin2\alpha$ and
$\cos2\alpha$, reducing the ambiguity in extracting $\alpha$
itself.  However, thousands of reconstructed events would be
needed to achieve the necessary sensitivity.  Whether either
the $B$ Factories at the $\Upsilon(4S)$ or the proposed
hadronic $B$ experiments BTeV and LHC--B can do such an
analysis remains an open and extremely important question.

\subsubsection{Direct CP violation in $B$ decays and the
extraction of $\gamma$}

There is a wide variety of proposals to measure the third
angle $\gamma$ of the Unitarity Triangle.  In constrast to
the other two angles, to measure $\gamma$ one must rely on
{\it direct\/} CP violation in $B$ decays.  This introduces
strong phases in an essential way, with two important
implications.  First, often they must be extracted or bounded
simultaneously with $\gamma$.  Second, the ultimate
sensitivity of a given construction to $\gamma$ typically
depends on strong phases which are not known beforehand.  Here
I will mention very briefly two classes of proposals for
measuring $\gamma$.

First, it is possible to extract $\gamma$ cleanly from rate
measurements in $B^\pm\to DK^\pm$ and $B_s\to D_sK^\pm$, but
both experiments are difficult.  The strategy in $B^\pm$ decay
is to find two final states $f_i$, each common to $D^0$ and
$\bar D{}^0$, and extract the strong and weak phases from the
four decays $B^\pm\to(D^0,\bar D{}^0)K^\pm\to
f_i\,K^\pm$.  The total branching ratios are at the level of
$10^{-7}$ and high precision is required, so it is not clear
whether the $B$ Factories at the $\Upsilon(4S)$ will have
sufficient statistics to complete such an
analysis~\cite{Harrison:1998yr}.  The second option, involving
$B_s$ decays, certainly must wait for BTeV or LHC--B.

The second class of methods is to extract $\gamma$ from
combinations of rates of $B\to K\pi$ modes.  While
there are many such approaches, all require additional
inputs of some kind~\cite{gammaproposals}.  The most common of
these are $SU(3)$ flavor symmetry and dynamical assumptions
concerning rescattering effects, the sizes of penguin diagrams,
or factorization of hadronic matrix elements. 
Unfortunately, there are too many such proposals to review
them usefully here.  Although none is as theoretically
clean as are the extractions from $B_{(s)}\to D_{(s)}K^\pm$
rates, analyses of this sort may be interesting if one treats
very conservatively the uncertainty from the additional
assumptions~\cite{gammaresults}.

\subsubsection{Measurement of $|V_{ub}|$}
\label{Vubsection}

Our current knowledge of
$|V_{ub}/V_{cb}|=\lambda\sqrt{\rho^2+\eta^2}$ comes from
semileptonic decays mediated by the tree level process $b\to
u\ell\,\bar\nu$.  Both inclusive $\bar B\to X_u\ell\,\bar\nu$
and exclusive $\bar B\to(\pi,\rho)\ell\,\bar\nu$ decays
have been used to extract this parameter.  All determinations
of $|V_{ub}|$ must contend with an enormous background from
$b\to c\ell\,\bar\nu$, which in the Standard Model is
approximately 100 times larger.  Inclusive analyses reject
charmed final states by imposing strict kinematic cuts, such as
$E_\ell>2.3\,$GeV or $M(X_u)<1.8\,$GeV. (The mass $M(X_u)$
is inferred by reconstructing the missing
neutrino.)  However, these cuts have an unfortunate
impact on the theoretical prediction of the decay
rate.  

The total rate is given at tree level by $\Gamma(\bar B\to
X_u\ell\,\bar\nu)={G_F^2 m_b^5/192\pi^3}\cdot
|V_{ub}|^2$, with radiative and nonperturbative corrections
which are well understood~\cite{Ligeti99}.  The largest
uncertainty comes from the value of $m_b$, which currently may
be determined with an error of approximately 100~MeV.  The
induced theoretical uncertainty in $|V_{ub}|$ is at the level
of 5\%.  As stringent kinematic cuts are applied, however, the
inclusive rate becomes much less inclusive, developing an
essential dependence on the momentum distribution of the $b$
quark inside the $B$ meson.  This is because in the presence
of the cuts one becomes sensitive to the {\it shape\/} of the
differential decay spectrum, not just to its integral.  On
scales of relative order $\Lambda_{\rm QCD}/m_b$, the spectrum
probes all moments of this $b$ quark ``wavefunction''.  Since
the wavefunction is not known from first principles and must
be modeled, this introduces an irreducible and uncontrolled
model dependence into the analysis.  The earliest
determinations of $|V_{ub}|$ from the endpoint region in
$E_\ell$ were completely polluted in this way, with results
varying by as much as a factor of two from model to model.

More recent determinations rely primarily on the cut
$M(X_u)<1.6\,$GeV, which keeps a significantly larger
fraction of the charmless final state phase space than does
a stringent cut on $E_\ell$ alone.  A recent LEP average
yields~\cite{LEPVub99}
\beqa
&|V_{ub}|=\left[4.05{}^{+0.39}_{-0.46}({\rm stat.})
{}^{+0.43}_{-0.51}(b\to c){}^{+0.23}_{-0.27}(b\to u)
\pm0.02(\tau_b)\pm0.16({\rm HQE})\right]\times10^{-3}\,,
&\nonumber
\eeqan
or approximately
$|V_{ub}/V_{cb}|=0.104^{+0.015}_{-0.018}$. 
While these analyses are experimentally very sophisticated,
they rely intensively on a two-parameter model of the $b$
quark wavefunction.  Essentially, in such a parameterization
all moments of the $b$ momentum distribution are correlated
with the first two nonzero ones, a constraint which is
unphysical.  Even if the two parameters are varied within
``reasonable'' ranges, it is doubtful that such a restrictive
choice of model captures reliably the true uncertainty in
$|V_{ub}|$ from our ignorance of the structure of the $B$
meson.  While the central value which is obtained in these
analyses is reasonable, the realistic theoretical error
which should be assigned is not yet well understood.

A recent analysis by CLEO of the exclusive decay
$\bar B\to\rho\ell\,\bar\nu$ yields~\cite{Behrens:1999vv}
\beqa
&|V_{ub}|=\left[3.25\pm0.14({\rm stat.})
{}^{+0.21}_{-0.29}({\rm syst.})\pm0.55({\rm
{\rm theory}})\right]\times10^{-3}\,, &\nonumber
\eeqan
or approximately $|V_{ub}/V_{cb}|=0.083^{+0.015}_{-0.016}$,
essentially consistent with the LEP
result.  In this case the reliance on
models is quite explicit, since one needs the hadronic form
factor $\langle\rho|\,\bar
u\gamma^\mu(1-\gamma^5)b\,|\bar B\rangle$ over the range of
momentum transfer to the leptons.  The CLEO
measurement relies on models based on QCD sum rules, which have
uncertainties which are hard to quantify.  Hence, just as in
the case of the LEP measurement, the quoted errors should not
be taken terribly seriously.  All of the current constraints
are consistent with
$|V_{ub}/V_{cb}|=0.090\pm0.025$, where I strongly prefer this
more conservative estimate of the theoretical errors. The
problem lies not in the experimental analyses, but in our
insufficient undertanding of hadron dynamics.

As is clear from Fig.~\ref{fig:constraints}, a reliable
measurement of $|V_{ub}|$ would provide an important
constraint in the $(\rho,\eta)$ plane.  What needs to be done
to improve the present situation?  For analyses based on
inclusive decays, the model dependence will be reduced only if
the kinematic cuts used to reduce the charm contamination can
be loosened.  The larger the fraction of the phase space for
$\bar B\to X_u\ell\,\bar\nu$ which is actually observed,
the smaller will be the sensitivity to the shape of the decay
spectrum.  For exclusive analyses, a more reliable
understanding of the form factors is needed.  Eventually, the
lattice should provide a good calculation of
$\langle\rho|\,\bar u\gamma^\mu(1-\gamma^5)b\,|\bar B\rangle$;
in this case, we are fortunate that the lattice computations
are most reliable in the region of large momentum transfer to
the leptons, where the experiments are also the most
sensitive.  Alternatively, one may use heavy quark and $SU(3)$
flavor symmetry to relate $\bar B\to\rho\ell\,\bar\nu$ to
$D\to \bar K{}^*\ell\nu$.  It will be crucial to understand the
leading corrections from symmetry breaking
effects~\cite{Ligeti:1996yz}.  Although the extraction of
$|V_{ub}|$ from $B\to\pi\ell\,\bar\nu$ is more complicated due
to the $B^*$ pole, dispersion relations possibly can
be used to control the hadronic form
factors~\cite{Boyd:1995tt}.  Finally, there have been
proposals to extract $|V_{ub}|$ from inclusive nonleptonic
decays~\cite{Chay:1999pa}.  This method is under excellent
theoretical control, but it remains to be seen whether it is
feasible experimentally.

\subsubsection{$B$ physics tests of the Standard Model}

The ultimate goal of a robust $K$ and $B$ physics program
is to probe the adequacy of the Standard Model
description of the LESM by comparing many independent
constraints in the $(\rho,\eta)$ plane.  Yet there also exist
simple strategies for looking for new physics in the $B$
system alone, which may prove fruitful long before the
complete program can be realized.  Here I will discuss briefly
two simple examples.

The first is to compare the CP asymmetries in $B\to
J/\psi\,K_S$ and $B\to\phi K_S$~\cite{Grossman:1997ke}.  In
the Standard Model, both of these measure the phase of
$B^0-\bar B{}^0$ mixing, since the decay amplitudes are real
to a good approximation.  For $B\to J/\psi\,K_S$, the decay is
dominated by a tree level process $b\to c\bar cs$, which is
large and therefore not so likely to receive significant
contributions from new physics.  By constrast, the $b\to s\bar
ss$ transition underlying $B\to\phi K_S$ is generated only by
penguin diagrams, so it carries a natural suppression of 
$\alpha_s/4\pi$.  Because of this suppression, it is possible
that loops with new heavy particles could contribute
significantly to $b\to s\bar ss$ but not to $b\to c\bar cs$. 
If these new loops carried nonzero phases, then the value of
``$\sin2\beta$'' extracted from $B\to\phi K_S$ would differ
from that extracted from $B\to J/\psi\,K_S$.  Such an
observation would be an unambiguous and exciting sign of new
physics.

The second example is a strategy for identifying new
contributions to the phase of $B^0-\bar B{}^0$
mixing~\cite{Grossman:1997dd}.  Note that if one compares
Eqs.~\leqn{lampsiks} and \leqn{lampipi} for the asymmetries in
$B\to J/\psi\,K_S$ and in $B\to\pi\pi$, the difference is independent of the mixing amplitude
$\M_B$ and sensitive to
\beq
2\arg\A_{\psi K_S}+\arg\M_K-2\arg\A_{\pi\pi}\,.
\eeq{diffgamma}
which is $2\gamma$ in the Standard Model.  If an isospin
analysis can be done in $B\to\pi\pi$ or $B\to\rho\pi$, the
difference of asymmetries
\leqn{diffgamma} can be combined with $\sqrt{\rho^2+\eta^2}$
from a measurement of $|V_{ub}/V_{cb}|$ to fix the point
$(\rho,\eta)$.  Once $(\rho,\eta)$ is identified, the angle
$\beta$ can be predicted, and thus one knows the phase
$\arg\M_B$ which should be extracted from $B\to J/\psi\,K_S$
if the mixing is given by Standard Model physics.  New
contributions to $B^0-\bar B{}^0$ mixing should show up in the
comparison, since there is no reason for them to have the
same nonzero phase as that predicted by $\vckm$.  Such an
analysis would be sensitive, for example, to mixing induced by
supersymmetric $\tilde t-\widetilde w$ box diagrams, as
expected in scenarios of electroweak baryogenesis in the
MSSM.

\section{A few concluding comments}

A rich program of experimental and theoretical flavor physics
awaits us in the first decade of the next century.  One of
the most delightful aspects of this field is the strong and
growing coupling between theory and experiment.  The
complications brought by hadronic physics force the two sides
of our community to work together to identify and develop
analyses which are both feasible experimentally and
controllable theoretically.

The most important goal in $K$ physics is to measure with 10\%
accuracy the branching ratios for the very rare decays
$K^-\to\pi^-\nu\bar\nu$ and $K_L\to\pi^0\nu\bar\nu$.  To do so
will require both extensive R\&D by physicists and continuing
support from the funding agencies.  But these experiments are
crucial, and we must commit to them.

In $B$ physics, a host of new experiments awaits us. 
Over the next decade we will have a wide variety of
measurements of CP violating asymmetries, branching
fractions, and mixing parameters in the $B$ and
$B_s$ systems.  There are important roles to be played by all
of the current and proposed detectors.  In the next few
years, the $B$ Factories at the $\Upsilon(4S)$ will take the
lead, along with the Tevatron experiments and HERA--B.  But in
the longer term it is crucial that there be a
dedicated $B$ experiment at a high energy
hadron collider, such as BTeV or LHC--B.  Such a detector
will be able to explore the $B_s$ system and observe some of
the rarest $B$ decays.  A luminosity upgrade
to one of the $\ee$ $B$ Factories also may be in our future.

Finally, a few philosophical points.  First, every
measurement of a new operator in the Low Energy Standard Model
is significant, {\it especially\/} if the measurement is
redundant from the point of view of constraining the Unitarity
Triangle.  Such comparisons are how we will find new physics
if it is there!  Second, one theoretically clean measurement is
worth many polluted ones, although even polluted ones will
become important as our understanding of hadronic physics
improves.  Third, the goal of this program is not simply to
measure $\rho$ and $\eta$, nor to constrain $\alpha$, $\beta$
and $\gamma$, nor to search for CP violation in the $B$
system.  Rather, it is to learn as much as possible about the
physics of flavor, and to probe robustly the Standard Model
description of flavor changing dynamics.  When new degrees of
freedom are seen at Fermilab Run II or at the LHC, their
virtual impact on ``low energy'' physics will be crucial in
helping us to understand in detail what we have actually
discovered.

\bigskip
This work was supported by the
National Science Foundation under grant PHY--9404057 and
National Young Investigator Award PHY--9457916, the
Department of Energy under Outstanding Junior Investigator
Award DE--FG02--94ER40869, and the Alfred P.~Sloan
Foundation.  A.F.~is a Cottrell Scholar of the Research
Corporation.

\vfill\eject

\def\Discussion{
\setlength{\parskip}{0.3cm}\setlength{\parindent}{0.0cm}
     \bigskip\bigskip      {\Large {\bf Discussion}} \bigskip}
\def\speaker#1{{\bf #1:}\ }

\Discussion

\speaker{Michael Peskin (SLAC)} 
This morning, Ron Poling showed
a fit for $\gamma$ from ratios of two-body exclusive $B$
branching ratios. {\rm [See also the CLEO
preprint~\cite{gammaresults}.]}  This result assumed that
factorization is perfect, and so it has to be taken with a
grain of salt.  But do you think it is possible that such a
determination of $\gamma$ could eventually be made
``theoretically clean'' in the sense that you used this term
in your talk?

\speaker{Falk}  
No, not in the restrictive sense in which I used the term. 
Analyses such as this one require severe assumptions about
hadronic dynamics, assumptions which are then partially
cross-checked within the analysis by examining the quality of
the fit.  But this ought not to be confused with actually
understanding the physics of hadrons!  The real test of an
analysis is whether it is theoretically clean enough that one
would be willing to rely on it for a definitive discovery
of physics beyond the Standard Model.  This is certainly not
true of an analysis based on factorization.  The natural
explanation of any discrepancy with the Standard Model
will be simply that the {\it ad hoc\/} assumptions that had
been made about hadronic dynamics were unjustified.

\speaker{B.~F.~L. Ward (University of Tennessee)}
I would like to make a comment on the penguin pollution in
$B\to\pi\pi$.  There is an interplay between the phases as
well as the size of the amplitude contributions (penguin and
non-penguin), and in general there may be regions of parameter
space in which $\sin2\alpha$ is measurable to some specificed
accuracy even though the penguin amplitude's magnitude is
large.  Thus, I think your comments were a bit too pessimistic.

\speaker{Falk}
I disagree.  It is true that hadronic physics may
conspire to make the asymmetry in $B\to\pi\pi$ equal to
$\sin2\alpha$ even in the presence of large penguin
contributions.  But this fact is of no use if the hadronic
matrix elements, both {\it magnitudes\/} and {\it phases,}
cannot be computed reliably from first principles.  If
$B\to\pi\pi$ is to provide any information on $\vckm$, we have
to know unambiguously the relationship between the observed
asymmetry and the operators of the LESM.  The problem with
penguin pollution is not simply that it may be large, it is
that we do not know what it is!

\speaker{Ken Peach (Rutherford Appleton Laboratory)}
When might the lattice be capable of making an accurate
prediction for either $\epsilon$ or $\epsilon'$?

\speaker{Falk} 
I am not an expert on these
issues, but I would argue that one will be able to use
lattice QCD reliably only when it is possible to go beyond the
quenched approximation.  This will start to come
in the next generation of lattice computations.  The
real question is, at what point will the lattice
theorists trust their calculations enough that they would
stick their necks out and rely on them for a controversial
claim such as that new physics is required to explain 
$\epsilon'$?  This will surely have to wait until unquenched
calculations are not only available, but well understood.

\end{document}